\documentstyle[12pt,aaspp4,epsf]{article} 

\begin{document} 

\title{
High Proper Motion Stars in the Vicinity of Sgr A*:  Evidence for a 
Supermassive Black Hole at the Center of Our Galaxy} 
\author{A. M. Ghez\altaffilmark{1,2}, B. L. Klein, M. Morris, \& E. E. Becklin}
\affil{Dept. of Physics and Astronomy, UCLA, Los Angeles, CA 90095-1562}
\authoremail{ghez,kleinb,morris,becklin@astro.ucla.edu}

\altaffiltext{1}{Sloan Fellow}
\altaffiltext{2}{Packard Fellow}

\begin{abstract}
Over a two year period, we have conducted a diffraction-limited imaging study 
at 2.2 $\mu m$ of the inner $6\tt'' \times $6$\tt''$ of the Galaxy's central
stellar cluster using the W. M. Keck 10-m telescope.
The K band images obtained in 1995 June, 1996 June, and 1997 May have the 
highest angular resolution obtained at near-infrared wavelengths from ground or
space ($\theta_{res}$ = 0\farcs 05 = 0.002 pc) and reveal a large population 
of faint stars.  We use an unbiased 
approach for identifying and selecting stars to be included in this proper 
motion study, which results in a sample of 90 stars with brightness ranging 
from  K = 9 to 17 mag and two-dimensional velocities as large as 1,400 $\pm$ 
100 km/sec.  Compared to earlier work (Eckart et al. 1997; Genzel et al. 1997),
the source confusion is reduced by a factor of 9, the number of stars with 
proper motion measurement in the central 25 $arcsec^2$ of our galaxy is 
doubled, and the accuracy of the velocity measurements in the central 1 
$arcsec^2$ is improved by a factor of 4.

The peaks of both the stellar surface density and the velocity dispersion are consistent
with the position of the unusual radio source and black hole candidate, 
Sgr A*, suggesting that Sgr A* is coincident ($\pm$ 0\farcs 1) with the 
dynamical center of the Galaxy.  As a function of distance from Sgr A*,
the velocity dispersion displays a falloff well fit by Keplerian motion
($\sigma_v \sim r^{-0.5 \pm 0.1}$) about a central dark mass of 
$2.6 (\pm 0.2) \times 10^6 M_{\odot}$ confined to a volume of at most
$10^{-6} pc^3$, consistent with earlier results.  
Although uncertainties in the measurements mathematically allow for the
matter to be distributed over this volume as a cluster, no realistic
cluster is physically tenable.  Thus, independent of the presence of Sgr A*, 
the large inferred central density of at least $10^{12} M_{\odot} / pc^3$, 
which exceeds the volume-averaged mass densities found at the center of any 
other galaxy, leads us to the conclusion that our Galaxy harbors a massive 
central black hole.

\end{abstract}


\section{Introduction}

Extrapolating from the idea that the highly energetic phenomena observed in 
very active galaxies are powered by massive central black holes, Lynden-Bell 
\& Rees (1971) suggested more than a quarter century ago that much less active 
galaxies such as our own Milky Way may also harbor massive, though possibly 
dormant, central black holes.  Early on, indirect support for a central black 
hole arose from the discovery of the unusual radio source Sgr A*;  its 
non-thermal spectrum (e.g., Serabyn et al. 1997, Beckert 
et al. 1996), compact size (Rogers et al. 1994) and lack of detected motion 
(Backer 1994) led researchers to associate it with the putative black hole.  
Definitive proof regarding the existence of a massive central black 
hole and its association with Sgr A*, however, lies in the assessment 
of the distribution of mass in the central few parsecs of the Galaxy.
If gravity is the dominant force, the motion of the stars
in the vicinity of the putative black hole 
reveals the mass interior to their orbital radius.  
Thus, objects located closest to the Galactic Center 
provide the strongest constraints on the black hole hypothesis.

To probe the inner region of the Galaxy, it is crucial to attain the highest
resolution possible.  However, turbulence in the Earth's atmosphere distorts 
astronomical images and typically limits the angular resolution of 
long-exposures to $\sim$0.5 - 1 arcsec, an order of magnitude worse than the 
theoretical limit for large ground-based telescopes.
With a distance of 8 kpc to the Galactic center (Reid 1993), traditional
long exposure observations are limited to estimating a central mass
constrained only to a volume of radius greater than or equal to $\sim$ 
0.1 pc (e.g., Lacy et al. 1980, McGinn et al. 1989, Haller et al. 1996;
Genzel et al. 1996).
In contrast to long exposures, short exposures such as the one shown 
in Figure 1a, although distorted by the atmosphere, preserve high spatial 
resolution 
information which can be used to recover diffraction limited images 
via a number of different techniques, such as the relatively simple
and straight forward method of ``Shift-and-Add" (Christou 1991).  
Eckart \& Genzel (1996, 1997) applied this method to data from the ESO 3-m NTT
and achieved a resolution of 0\farcs 15 in the first proper motion
study of the central stellar cluster.  
This technique applied to data obtained from the 
W. M. Keck 10-meter telescope provides a unique opportunity to study the Galaxy center
at an unprecedented resolution of 0\farcs 05.  

Here we report the initial results of our proper motion 
study of the Galaxy's central stellar cluster.  Within our 
$6\tt'' \times $6$\tt''$ field of view, the motions of 90 stars are tracked over two years.
With two-dimensional velocities as high as 1,400 km/sec,
these stars imply a central mass of $2.6 \pm 0.2 \times 10^6 M_{\odot}$ 
interior to a radius of $\sim$ 0.015 pc, or densities in excess of
$10^{12} M_{\odot} / pc^3$.  This volume-averaged mass density exceeds
that inferred so far for the center of any other galaxy.  The high mass to light 
ratio and high density lead us to conclude that our Galaxy harbors a massive 
central black hole.

\section{Observations \& Data Analysis}

We observed the Galaxy's central cluster with the Keck I
facility near-infrared camera (NIRC; Matthews \& Soifer 1993) on 
the nights of 1995 June 10 - 12, 1996 June 26 - 27, and 1997 May 13
(UT).  In order to obtain the highest angular resolution with the maximum
sensitivity to stars in this region, measurements were made in the photometric 
K bandpass ($\lambda_o$=2.2$\mu m$  $\Delta \lambda$=0.4 $\mu m$) with a 
magnified plate scale of 0\farcs 0203/pixel, providing a 5\farcs 12 $\times$ 
5\farcs 12 field of view (Matthews et al. 1996).  Each year, 3,000 - 6,000 
images, with an individual exposure time of 0.13 sec, were obtained in sets of 
100.  These images were primarily centered on the location of Sgr A*, but a few 
hundred offset frames were also obtained in order to generate a mosaic of a larger 
region that includes 2 SiO maser sources - IRS 7 and 10ee (Menten et al. 1997)
- thereby linking the infrared and radio reference frames, and securing the 
position of Sgr A* to $\pm$ 10 mas (1$\sigma$).
 
Conversion of the raw data into diffraction-limited images
proceeds in two steps: (1) calibration of the individual frames and (2)
creation of the shift-and-add maps.  First, the standard image analysis
steps of sky subtraction, flat fielding, and bad pixel correction remove the 
majority of the measurement artifacts from the individual frames,
which are known as specklegrams.  
A few additional operations occur at this initial stage of analysis;
a small area ($\sim$ 12\% of the field of view), which is contaminated by the 
unmagnified field (cf. Matthews et al. 1996), is masked off, the camera's 
distortion is accounted for (see Appendix for details),
each pixel is expanded into 2$\times$2 pixels with equal flux,
in order to prevent degradation of the image quality when
shifting the specklegrams by fractions of a pixel, and, finally, the specklegrams are rotated through the 
parallactic angle specified in the header such that they have a common 
orientation\footnote{In these experiments, the image rotator 
was turned off such that the field orientation on the detector changed 
throughout the night.  This procedure minimizes the systematic effects of possible flat 
fielding imperfections and image distortion, and maximizes the sky coverage 
in the final shift-and-add maps.}. 
After this first stage of data processing the specklegrams are
free of camera artifacts and are dominated by the short
exposure structure of the 4 brightest stars, IRS 16 NE, IRS 16 NW, IRS 16 C,
and IRS 16 SW (K $\sim$ 9-10 mag; see Figure 1a). 

In the second stage of data reduction, the frames are combined to 
form shift-and-add maps.  Each star in the individual exposures is distorted 
by the atmosphere in the same way, indicating that the field of view is well 
within the near infrared isoplanatic patch, and has a pattern that is 
dominated by one bright diffraction-limited spot (or speckle).  
By adding together the specklegrams, shifted to align the brightest speckle of
a reference source (IRS 16 C), we generate a shift-and-add image 
with a point spread function composed of a diffraction-limited core
containing $\sim$10\% of the light on top of a broad seeing halo   
(see Figures 1 \& 2).  Airy rings around the cores of bright stars
indicate that the diffraction limit has been truly achieved.  
The atmospheric conditions during each run  
are evaluated on the basis of the halo FWHM, which is
0\farcs 5 in 1995, 0\farcs 8 in 1996, and 0\farcs 4 in 1997.
Due to field rotation and differences in the centering
of individual frames, the field of views of the final shift-and-add images 
are larger than the original frames, with a $6\tt'' \times 6\tt''$ region
containing all the stars reported here.

\section{Results}

\subsection{Stellar Census}

The raw shift-and-add maps reveal a large population of faint stars
against the seeing halos of the bright IRS 16 stars.  The primary objective 
in the design of the analysis presented here is to generate an {\em unbiased} 
sample, which is defined independently of the position of Sgr A*.  Stars are 
identified in each image by applying a ``match filter," generated by 
cross-correlating the raw shift-and-add image with the diffraction-limited 
core (out to r $=$ 0\farcs 06) of IRS 16 NE, the brightest isolated point 
source in the field. The values in the correlation map range from 1 for 
perfectly correlated regions to -1 for perfectly anti-correlated regions;  
strong peaks, such as the ones shown in Figure 1d, 
mark the locations of stars.  Each peak in the correlation maps exceeding 0.2 
is flagged as a {\it potential} star, and the star's brightness is estimated 
by aperture photometry at that location in the shift-and-add map (for details 
see footnote to Table 1).
{\it Definite} detections are marked for cases which have both a correlation
peak exceeding 0.7 and more than 1,000 frames contributing to that
location in the final shift-and-add map.  
Roughly 80 definite detections are made in each map.
The final population of stars in each year's map that is considered in this 
proper motion study is composed of
the set of definite detections for a given year and the matches
to the definite detections from the other 
two years, which are searched for among the possible detections  
within a 0\farcs 07 radius. 
If more than one match is found, the closest star must be the 
most similar in brightness or the star is eliminated from the
sample due to source confusion.
Table 1 lists the 90 stars that are thus identified in all three measurements 
and comprise the proper motion sample;
half of these stars are definite detections in all three years, whereas 
the other half have definite detections in only one or two years.
The latter group arises from (1) the poorer quality of the 1996 map,
which causes some stars' correlation values to drop below the definite detection threshold
that year, but nonetheless satisfy the potential star criteria and are matched
to definite detections in the other years, and (2) differences in map centers, 
which results in some stars near the peripheries 
having more than 1000 contributing frames in at least one map but not in
all three.  
The stars in this proper motion sample have 
brightness ranging from K = 9 to 17 mag, 
or equivalently $M_K$ = -8.5 to -0.5 mag assuming $A_K$ = 3 mag,
with typical relative positional 
uncertainties of $\sim$ 0\farcs 002 for sources brighter than 15 mag.
The naming convetion we have adopted is described in Table 1.
\footnote{The positional uncertainties are based on the 
variance of the stars' locations in three independent and equivalent seeing submaps from each year.}

The distribution of stars in the proper motion sample shows a clear central 
concentration within 0\farcs 2 $\pm$ 0\farcs 1 of the position of 
Sgr A* (Menten et al, 1997).  It should be noted, however, that our 
sample of 90 stars itself is neither complete nor uniform.  Many stars in the 
image are missed using the fairly conservative approach and selection
criteria described above, which tends to omit
close doubles as their correlation values are significantly lower
than isolated stars - the smallest separation between
two stars in the final sample is 0\farcs 08.  
Furthermore, the seeing halos cause the sensitivity to faint sources to
grow with distance from bright
sources out to 1\farcs 5 where it is then more or less constant at 
a brightness $\sim$7 mag fainter than the bright source.
Nonetheless, with a peak value of $\sim$ 15 $stars/arcsec^2$, 
the stellar surface density for the proper motion sample appears
to manifest a distinct cluster, the Sgr A* (IR) cluster,
of stars fainter than K = 14 mag and an apparent core radius of 0\farcs 3 (0.01 pc).  

\subsection{Stellar Kinematics}

Since measuring the motions of stars requires a common coordinate system, 
the coordinate systems for the three measurements are aligned
by minimizing the net displacements
of {\em all} the stars in the proper motion sample, 
allowing for translation and rotation between the three epochs.  
Figure 3 shows the transformed positions for stars within a 1 $arcsec^2$
region centered on the nominal position of Sgr A*, where motion for several
of the stars can be easily seen.
Formal two-dimensional velocities are derived by fitting lines
to the positions as a function of time, weighted by the
positional uncertainties,
for all stars in the sample, of which 30 have significant motion
(v/$\sigma_v>$4) over the two year baseline of this study
(see Table 1).  In all cases, the velocity vectors measured between 1995-1996 
and 1996-1997 agree to within 4$\sigma$.  As a consequence of 
our approach to defining the proper motion sample the uncertainties
are {\em independent} of the overall location and depend primarily on the 
source brightness, with typical uncertainties of 50 km/sec for the brighter 
stars (K $<$ 13 mag), 100 km/sec for the intermediate brightness stars
(13 mag $<$ K $<$ 15 mag), and 200 km/sec for the faintest
stars (15 mag $<$ K $<$ 17 mag).  Eleven stars have proper motions
exceeding 500 km/sec, with the largest velocity reaching
1,400 $\pm$ 100 km/sec, 0.5\% the speed of light, for a K = 15 mag source
0\farcs 1 west of Sgr A*.

The distribution of velocities is nonuniform, with the
highest velocity sources clustered toward the field center (see Figure 4). 
Seven of the 11 sources with velocities exceeding 500 km/sec appear to be 
members of the central Sgr A* stellar cluster, with projected distances 
from the nominal position of Sgr A* of less than 0\farcs 8. 
A map of the two-dimensional velocity dispersion,  
obtained by binning the velocities over 1\farcs 5 $\times$ 1\farcs 5 
regions, shows a prominent peak of 670 km/sec located just 
0\farcs 1 East and 0\farcs 1 North of Sgr A*.  
The coincidence between the nominal position of Sgr A*, the 
peak of the stellar surface density, and the peak of the velocity 
dispersion, suggests that Sgr A* is indeed at the dynamical
center of our galaxy.  We therefore take the position of Sgr A* as the center
for the analysis that follows. 

With more than one dimension of the velocity vectors measured, we can
test for isotropy, in which case the velocity dispersions should be identical 
in all three dimensions.  In the plane of the sky, the projected radially
directed ($\sigma_{\parallel}$) and the projected tangential ($\sigma_{\perp}$) components of the velocity 
dispersion are similar at all radii covered by this study, with 
$\sigma_{\parallel} / \sigma_{\perp}$ = 1.15 $\pm$ 0.1.  
Line of sight velocity measurements for individual stars
have only been obtained for the largest radii covered here,
$\sim$0.1 pc; nonetheless, 
the velocity dispersion in the line of sight ($\sigma_z$) and that
of either dimension in the plane of the sky 
is in agreement, with $\sigma_z / <\sigma_{\parallel}, \sigma_{\perp}>$ = 
1.1 $\pm$ 0.2, as was shown by Genzel et al. (1996).  
The apparent isotropy of the stellar velocity field
suggests that the stars are moving under the influence of
a spherical potential.  

The apparent isotropy justifies averaging the two components of 
dispersion in the plane of the sky to examine only a one-dimensional velocity 
dispersion as a function of projected radius (Figure 5).  The velocity 
dispersion at small radii
is clearly much higher than the 50 km/sec dispersion observed at larger 
radii.  Furthermore, fitting these data to a power law, 
$\sigma_v (r) \sim r^{\alpha}$, results in best fit $\alpha$ of -0.53 $\pm$ 0.1,
an excellent match with that expected from Keplerian orbits ($\alpha$ = -0.5). 
This behavior suggests that {\em the stars' motions are dominated by the 
gravitational force of a large central mass 
confined to a radius less than the smallest radial bin - 0.015 pc.}
It also suggests that this central mass completely dominates the mass 
distribution out to at least the radius of the outermost bin, 0.1 pc.

\subsection{Comparison with Other Proper Motion Measurements}

Of the 90 stars presented here, roughly half -  44 stars - have velocities also 
reported by Eckart and Genzel 
(EG; Eckart \& Genzel 1996,1997; Genzel et al. 1997) 
that differ by no more than 4 $\sigma$ from those in Table 1; 
the largest differences arise in the central
$\sim 1 arcsec^2$ where the source confusion is the largest and the
increase in angular resolution has the most impact.
Over the $\sim$ 30 $arcsec^2$ area covered by both EG and this study,
only 3 of EG's 47 stars are not included in our proper motion sample. 
One of these - S5 - is resolved as a close, equal-brightness double, causing 
it to fail our selection criteria, although it is clearly seen in the
shift-and-add images.  
The other two - W15 and S3 - are selected in at least one of our epochs
but not all in three. The former is very close to the edge 
of the field of view and small shifts in the field center cause
it to fall below our detection limit in 1996 and 1997, and the 
latter is coincident with the Sgr A* star cluster where 
the source confusion is greatest.
S3 is identified as a definite detection in 1995, but is dropped from the
proper motion sample due to ambiguity when the search for
possible matches in the 1996 and 1997 maps reveal several possible solutions 
that do not satisfy the condition that the closest source has the most similar
brightness.  Although one of these solutions does produce constant velocity
(880 $\pm$ 200 km/sec at PA = 320$^o$), it and all the other solutions with
a search radius of 0\farcs 07 require the source to fade by 1.5 mag between 
1995 and 1996.  More frequent observations at as high a spatial resolution 
as possible are clearly 
needed to unambiguously track both the motion and variability of sources in 
this high density region. 
The proper motions for the remaining 46 stars listed in Table 1, which have magnitudes 
ranging from 12 to 17, 
are the first reported values for these stars; they {\em double} the 
number of stars with proper motion measurements in this region. 

In addition to the proper motion stars, Genzel et al. (1997) also report S12 as
a variable star and as a possible infrared-counterpart to Sgr A*.  
This source is detected as a possible source in our 1996 and 1997 data sets, 
but does not meet the proper motion selection criteria. 
The proper motion derived from the 1996 and 1997 measurements, however, is inconsistent
with the proper motion limits derived for Sgr A* at radio wavelengths 
(Backer 1994).

\section{The Central Dark Mass}

The two-dimensional positions and velocities measured for stars in the 
inner 6$\tt'' \times $6$\tt''$ (0.23 pc $\times$ 0.23 pc) 
provide excellent constraints on the distribution of matter at the center
of the Galaxy.   In principle, if all six components of the position
and velocity vectors could be observed, each star would yield an estimate 
of the mass enclosed within its radius.  With
the two-dimensional projections, the individual stars provide only lower
limits on the enclosed mass, $M_{min}$, under the assumption that the stars are 
gravitationally bound, in which case 
\[ M_{min} =  \frac{v^2 R}{2G}. \]
Every star imposes a minimum mass that exceeds the enclosed mass
of luminous matter extrapolated from the power law relationship derived at
larger radii by Genzel et al. (1996).
Considering only stars with $v/\sigma_v \ge 4$,
the minimum enclosed mass estimates reach values
of $2-3 \times 10^6 M_{\odot}$, with the apparent members
of the Sgr A* cluster having $M_{min}$'s ranging from 
$0.2-1 \times 10^6 M_{\odot}$ (see Figure 6).
Thus the stars appear to be moving under the influence of a gravitational
potential generated by at least a few million solar masses of dark matter.

Projected mass estimators analyze stars grouped in concentric annuli around 
the dynamical center to account for projection effects and produce
estimates of the true enclosed mass.  The well-known and frequently used 
virial mass estimator, $M_{virial}$, has the form 
\[ M_{virial} = \frac{3 \pi }{2G} <v^2>  / <1/R>. \]
Applied to the Galactic center data set, this mass estimator
suggests that 
$2.5 \pm 0.2 \times 10^6 M_{\odot}$ of dark matter is located
predominantly, if not exclusively, at radii smaller than 0.015 pc.
Bahcall \& Tremaine (1981), however, pointed out that the virial mass
is a biased, inefficient, and, in some cases, inconsistent mass
estimator; they proposed a set of new projected mass estimators for
a tracer population moving under the influence of a central potential,
$M_{BT}$.  For the case of isotropic orbits this estimator is given by 
\[ M_{BT} = \frac{16}{\pi G}  <v^2 R>. \]
The assumption of a dominating compact central mass is well
justified, given the Keplerian fall-off of the velocity dispersion
as a function of radius (Figure 5).  
The validity of this assumption allows one to use the projected
mass estimator with only minor caveats about the finite sampling
volume (Haller \& Melia 1996), implying a possible correction on 
the order of 10\% (which we do not apply).

For the Galactic center data set the two methods produce very
similar results, with the $M_{BT}$ values being only slightly larger,
$2.6 \pm 0.2 \times 10^6 M_{\odot}$.  
These values agree very well with those obtained over similar radial distances
by Genzel et al. (1997) and match well with those obtained at larger radii
(see Figure 7).
Overall, the enclosed mass results suggest, consistent with both the functional
form of the velocity 
dispersion vs. radius and the minimum mass estimates, 
that the majority of stars observed are moving in a potential 
dominated by $2.6 \times 10^6 M_{\odot}$ of matter contained 
within 0.015 pc of Sgr A*. 
Since the total luminosity within 0.015 pc of Sgr A* observed in our maps
is a meager $L_K$ of 40 $L_{\odot}$, the implied mass to light ratio is
$M/L_K \sim 6 \times 10^5$.  As the K-band covers only a small range of wavelengths,
it is useful to compare this value to that observed for the Sun, which
has a $M/L_K$ of $\sim$ 40.
Given the high mass to light ratio observed, the central mass concentration is certainly 
composed primarily of dark matter. 

Strong constraints on the distribution of central dark
matter arise from the enclosed mass measurements.  
Confining the density distribution of dark matter 
to radii smaller than 0.015 pc implies
a minimum density of $10^{12} M_{\odot}pc^{-3}$,
surpassing the volume averaged mass densities
inferred for dark matter at the center of any other galaxy 
by at least two orders of magnitude.  
One intriguing possibility is that the 
dark compact object we are observing is 
a single supermassive black hole, as has been inferred for several
other galaxies such as M87 (Ford et al. 1994; Harms et al. 1994) and 
NGC 4258 (Greenhill et al. 1995; Myoshi et al. 1995).
This would be a unique solution  if the minimum radius of the enclosed mass 
measurements corresponded to the Schwarzchild radius for a 
$2.6 \times 10^6 M_{\odot}$ black hole ($R_{sh} = 2GM_{bh}/c^2$,
$R_{sh, 2.6 \times 10^6 M_{\odot}} = 2.5 \times 10^{-7} pc = 11 R_{\odot}$),
in which case the central ``density" would be 
$\rho_{o,bh}  = 4 \times 10^{25} M_{\odot}pc^{-3}$;
however the minimum radius is still a factor
of 40,000 larger than $R_{sh}$ and thus other scenarios still need to be explored. 

One alternative to the single black hole scenario is 
a cluster of dark matter in the form of 
stellar remnants, brown dwarfs, or even elementary particles.  
In general, astrophysical clusters can often be approximated by a
Plummer model,
$ \rho(r) = \rho_o \left( 1 + \frac{r^2}{{r_c}^2} \right)^{- \alpha / 2} $,
which requires the specification of two parameters in addition to the 
central density: a characteristic size scale, $r_c$,
and the power law, $\alpha$.  Although an $\alpha$ of 2 holds
for the visible stellar cluster with $\rho_o = 4 \times 10^6 M_{\odot}/pc^3$
and $r_c$ = 0.2 pc, such a profile produces an enclosed mass
which increases linearly with radius, much steeper 
than that observed within the central 0.2 pc.  To match the observed
flat enclosed mass as a function of radius with a pure cluster 
model requires $\alpha$ to be at least 3 and $r_c$ to be very small.   
Since astrophysical systems have
been observed with $\alpha$'s as large as 5,
we explored the viability of clusters with $\alpha$ ranging from 3 to 5.
Mathematically, dark cluster models can be made to fit the observed data; 
$\alpha$ of 5 
requires a $r_c$ of 0.01 pc and $\rho_o$ of $6 \times 10^{11} M_{\odot}/pc^3$,
$\alpha$ of 4 is fit by $r_c$ of 0.005 pc and $\rho_o$ of $2 \times 10^{12} 
M_{\odot}/pc^3$, and $\alpha$ of 3 demands $r_c$ of 0.00002 pc and $\rho_o$ 
of $7 \times 10^{18} M_{\odot}/pc^3$ (see Figure 7).  Physically, however, 
such dark cluster models are highly improbable (cf. Maoz 1995, 1998).  A 
viable cluster must have both evaporation and collision timescales 
greater than the lifetime of the Galaxy $\sim$ 10 Gyr.  Clusters of 
objects having any single mass greater than 0.02 $M_{\odot}$
have evaporation timescales shorter than the age 
of the Galaxy, ruling them out from 
consideration.  Among possible nonluminous cluster members with masses less 
than 0.02 $M_{\odot}$, brown dwarfs, and very low mass objects 
with cosmic composition are ruled out by their short collisional timescales, 
which are at most $10^7$ years. What has not been eliminated with timescale 
considerations alone are clusters of elementary particles and very low 
mass (M $<$ 0.02 $M_{\odot}$) black holes, however such clusters are 
theoretically unmotivated.  Thus, the observed mass distribution is 
not likely to be due to a pure cluster of dark objects.  

Another alternative is for only a fraction of the mass to be in a central 
black hole with the remaining mass contained in a cluster of dark objects as 
might be found in a post core-collapsed cluster. 
Fitting the measured enclosed mass as a function of radius with
a black hole plus an $\alpha \sim 2$ cluster model, we find that
only 1\% of the total mass interior to 0.015 pc can be in the cluster
due to rapid rise of the mass enclosed by an $\alpha \sim 2$ cluster.
Although larger $\alpha$ clusters relax this criterion, 
$\alpha \sim 2$ is the expected form for a cluster surrounding
a black hole (e.g., Binney \& Tremaine 1987). 
Thus the dynamical evidence, independent of the presence
of Sgr A*, leads us to the conclusion that our Galaxy harbors a 
$2.6 \times 10^6 M_{\odot}$ black hole.

Our Galaxy was neither the first nor an obvious candidate for a central
supermassive black hole; however in the million solar mass range, it, 
along with NGC 4258, has become one of 
the strongest cases for a black hole.  
The significance of a central black hole in our normal inactive Galaxy
is the implication that massive black holes might be found at the centers of 
almost all galaxies.

\acknowledgements

We thank the staff of the W. M. Keck Observatories, especially 
Joel Aycock, Teressa Chelminiak, Al Conrad, Bob Goodrich, Wendy 
Harrison, Check Sorenson, and Wayne Wack.
Support for this work was provided by NSF, the Packard Foundation, and
the Sloan Foundation.  The authors are grateful to M. Jura, R. White,
and J. Patience for helpful comments.  AMG is grateful for the support
for this project received from the National Science Foundation Young 
Investigator Program, the Packard Foundation, and the Sloan Foundation.
The data presented herein were obtained at the W. M. Keck Observatory,
which is operated as a scientific partnership among the University of 
California, the California Institute of Technology, and the
National Aeronautics Space Administration.  The Observatory was
made possible by the generous financial support of
the W. M. Keck Foundation.

\appendix

\section{NIRC Off-Axis Distortion}

NIRC, which sits $\sim$50$\tt''$ away
from the telescope's optical axis, has a known off-axis distortion which is
corrected for by applying the following relations from Gleckler (1995) for
the distorted ray positions $(x_D,y_D)$ relative to the true ray
positions $(x,y)$: $x_D = x+Bx+Cxy+Exr^2 $ and
$y_D = y+A+3By+Cy^2+(D+Ey)r^2$
where $r^2=x^2+y^2$, A = -0.00001708, B = -0.0002197, C = -0.003553, D = 0.001778,
and E = 0.00006560. 
The corrected intensity value at each (x,y) is found by bilinear interpolation
of the original image at the corresponding
distorted ray position, $(x_D,y_D)$.  This procedure reduces
the pixel scale variations from 2\% to less than 0.5\% over the field of view  
for a single frame.  In the final shift-and-add maps, averages
over different camera orientations further minimize any residual
distortions.

\pagebreak

\figcaption{
(a) One of the many short exposures ($t_{exp}$ = 0.13 sec) 
obtained on the central stellar cluster.   Each star has the same speckle
pattern, which is dominated by one bright speckle.  
(b) By registering all the individual frames with respect to IRS 16 C, 
a shift-and-add image, such as the one shown for 1995, is constructed.  
This image has 8 mag of dynamic range and is strecthed to show the brighter stars 
in the image.   (c) A fainter group of star are clustered within 0\farcs 5 
of Sgr A*.  A high pass filter has been used, for display purposes only, 
to eliminate the seeing halos of the neighboring IRS 16 stars.  
(d)  The correlation of the shift-and-add image, shown in (b), 
and the diffraction-limited core of the point spread function, scaled
to show the stars that are marked as definite detections.  
The location of stars are easily identified in this crowded field.  
All images are oriented such that North is up and East is to the left.}

\figcaption{(a) The point spread function for the 1995 June shift-and-add image
as measured from the radial profile of IRS 16 NE.   The diffraction limited core,
which contains $\sim$10\% of the flux, is built up from the brightes speckle in each
contributed frame, while the seeing halo results from the fainter surrounding speckles.
Airy rings encircle each core indicating that the diffraction-limit has truly
been achieved
(b) A gaussian fit to the seeing halo has been subtracted from the data
(dotted line) to emphasize the Airy ring contribution and the theoretical point
spread function of Keck (solid line) is plotted for comparison.}

\figcaption{The measured positions of stars a 1 $arcsec^2$ centered on 
the position of Sgr A* (starred point, which depicts the 
location of this radio source). 
Significant velocities, which reach 1400 $\pm$ 100 km/sec, are easily detected in this region.
Each year's measurement is represented by a different symbol: 1995 by triangles, 
1996 by squares, and 1997 by circles. }

\figcaption{The positions of the 90 stars
that were unambiguously detected in all three years are displayed with
pointsizes scaled to the stars' velocities.  Proper motion measurements with SNR 
of at least 4 are plotted as filled points.  A clear increase in the velocities
is visible at the field center, where stars reach velocities of 1,400 km/sec.}

\figcaption{The projected stellar velocity dispersion as a function of
projected distance from Sgr A* is consistent with Keplarian motion,
which implies the gravitational field is dominated by mass within 0.1 pc.}

\figcaption{The minimum enclosed mass as a function of projected radius
inferred from the 30 stars with $v/\sigma_v \ge 4$.  The solid
curve in the lower right corner is an extrapolation of the enclosed
luminous matter curve from Genzel et al. (1996).  Each star suggests
the presence of a central dark mass of roughly a million $M_{\odot}$. }

\figcaption{
The enclosed mass as a function of projected
distance from Sgr A* are shown for the results of
this study (7 filled circles), EG's proper motion study
(4 unfilled circles), Genzel et al. (1996) radial velocity
study (13 unfilled squares), and Guesten et al. (1987) measurement
of the rotating gas disk (2 unfilled triangles).  From 0.1
to 0.015 pc the enclosed mass appears to be constant 
with a value of $2.6 \times 10^6 M_{\odot}$.
Mathematically, power law dark clusters with $\alpha \ge 3 $ fit
the observed distributions, however they are not physically tenable
(see text).  The high density of the central dark mass, which 
exceeds $10^{12} M_{\odot}/pc^3$, is indicative of a single
supermassive black hole.}

\pagebreak

\begin{deluxetable}{llrrrrrrc}
\tablenum{1}
\footnotesize
\tablecaption{Keck Proper Motions Measurements}
\tablehead{\colhead{Star ID\tablenotemark{a}} &
	   \colhead{Other} & 
           \colhead{K \tablenotemark{b,c}} &
           \colhead{R \tablenotemark{b}} &
           \colhead{$\Delta RA$ \tablenotemark{b}} &
           \colhead{$\Delta DEC$ \tablenotemark{b}} &
           \colhead{$V_{ra}$} &
           \colhead{$V_{dec}$} &
	   \colhead{Notes} \nl
	   \colhead{} &
	   \colhead{} &
	   \colhead{(mag)} &
	   \colhead{(arcsec)} &
	   \colhead{(arcsec)} &
	   \colhead{(arcsec)} &
	   \colhead{(km/sec)} &
	   \colhead{(km/sec)} &
	   \colhead{} }
\startdata
S0-1 & S1	& 14.9 & 0.114 & -0.107 & 0.039 & 470 $\pm$ 130 & -1330 $\pm$ 140 & 1 \nl 
S0-2 & S2 	& 14.1 & 0.151 & 0.007 & 0.151 & -290 $\pm$ 110 & -500 $\pm$ 50 & 1 \nl 
S0-3 & S4 	& 14.7 & 0.218 & 0.198 & 0.090 & 495 $\pm$ 60 & 300 $\pm$ 50 & 1 \nl 
S0-4 & S8 	& 14.5 & 0.338 & 0.300 & -0.154 & 720 $\pm$ 80 & -530 $\pm$ 110 & 1 \nl 
S0-5 & S9 	& 15.1 & 0.350 & 0.212 & -0.279 & 120 $\pm$ 140 & -630 $\pm$ 250 & 1 \nl 
S0-6 & S10	& 14.2 & 0.440 & 0.148 & -0.414 & -400 $\pm$ 100 & 230 $\pm$ 100 & 1 \nl 
S0-7 & S6	& 14.9 & 0.460 & 0.453 & 0.077 & 480 $\pm$ 170 & 120 $\pm$ 130 & 1 \nl 
S0-8 & -	& 16.1 & 0.468 & -0.279 & 0.376 & 260 $\pm$ 260 & -850 $\pm$ 260 & 3 \nl 
S0-9 & S11	& 14.1 & 0.553 & 0.150 & -0.532 & 200 $\pm$ 60 & -80 $\pm$ 140 & 1 \nl 
S0-10 & -	& 15.2 & 0.578 & 0.300 & -0.494 & 120 $\pm$ 80 & -50 $\pm$ 160 & 3 \nl 
S0-11 & S7	& 15.5 & 0.598 & 0.597 & -0.023 & -130 $\pm$ 70 & -220 $\pm$ 130 & 1 \nl 
S0-12 & W6	& 14.4 & 0.625 & -0.521 & 0.345 & -100 $\pm$ 80 & 210 $\pm$ 30 & 1 \nl 
S0-13 & -	& 13.5 & 0.752 & 0.586 & -0.471 & -90 $\pm$ 60 & 250 $\pm$ 50 & 3 \nl 
S0-14 & W9	& 13.8 & 0.794 & -0.740 & -0.287 & 20 $\pm$ 90 & 50 $\pm$ 90 & 1 \nl 
S0-15 & W5	&13.8 & 0.911 & -0.852 & 0.321 & -310 $\pm$ 60 & -310 $\pm$ 130 & 1 \nl 
S1-1 & -	& 13.3 & 1.006 & 1.006 & 0.018 & 200 $\pm$ 90 & 60 $\pm$ 90 & 3 \nl 
S1-2 & -	& 14.9 & 1.014 & -0.055 & -1.013 & 510 $\pm$ 110 & 90 $\pm$ 170 & 3 \nl 
S1-3 & -	& 12.3 & 1.016 & 0.566 & 0.844 & -480 $\pm$ 50 & 150 $\pm$ 70 & 2 \nl 
S1-4 & -	& 12.9 & 1.045 & 0.772 & -0.705 & 410 $\pm$ 80 & 50 $\pm$ 100 & 3 \nl 
S1-5 & -	& 12.8 & 1.053 & 0.433 & -0.960 & -300 $\pm$ 70 & 230 $\pm$ 80 & 3 \nl 
S1-6 & -	& 16.0 & 1.061 & -0.822 & 0.672 & -600 $\pm$ 210 & 190 $\pm$ 270 & 3 \nl 
S1-7 & -	& 16.9 & 1.063 & -0.943 & -0.491 & -110 $\pm$ 250 & -200 $\pm$ 180 & 3 \nl 
S1-8 & W11	& 14.2 & 1.077 & -0.659 & -0.852 & 160 $\pm$ 60 & -240 $\pm$ 90 & 1 \nl 
S1-9 & 16 NW	& 10.1 & 1.190 & 0.027 & 1.190 & 310 $\pm$ 60 & 380 $\pm$ 110 & 2 \nl 
S1-10 & W8	& 15.0 & 1.204 & -1.199 & -0.114 & 120 $\pm$ 170 & 330 $\pm$ 60 & 1 \nl 
S1-11 & 16 C	& 10.0 & 1.294 & 1.218 & 0.436 & -370 $\pm$ 60 & 380 $\pm$ 40 & 2 \nl 
S1-12 & W13	& 13.7 & 1.316 & -0.852 & -1.003 & 220 $\pm$ 70 & 20 $\pm$ 60 & 1 \nl 
S1-13 & W12	& 14.2 & 1.357 & -1.032 & -0.880 & -410 $\pm$ 120 & -260 $\pm$ 140 & 1 \nl 
S1-14 & W10	& 12.9 & 1.371 & -1.340 & -0.287 & -20 $\pm$ 70 & -100 $\pm$ 60 & 1 \nl 
S1-15 & W4	& 14.3 & 1.410 & -1.311 & 0.519 & 30 $\pm$ 160 & -30 $\pm$ 90 & 1 \nl 
S1-16 & 16 SW	& 10.2 & 1.442 & 1.055 & -0.983 & 170 $\pm$ 60 & 150 $\pm$ 40 & 2 \nl 
S1-17 & -	& 12.4 & 1.549 & 0.545 & -1.450 & -200 $\pm$ 50 & -80 $\pm$ 50 & 3 \nl 
S1-18 & -	& 15.4 & 1.599 & -0.638 & 1.466 & -220 $\pm$ 190 & 90 $\pm$ 120 & 3 \nl 
S1-19 & -	& 13.8 & 1.630 & 0.382 & -1.584 & 20 $\pm$ 50 & -60 $\pm$ 120 & 3 \nl 
S1-20 & -	& 12.7 & 1.641 & 0.414 & 1.588 & 240 $\pm$ 80 & 160 $\pm$ 50 & 2 \nl 
S1-21 & W7	& 13.7 & 1.653 & -1.648 & 0.137 & 160 $\pm$ 120 & -200 $\pm$ 60 & 1 \nl 
S1-22 & W14	& 12.8 & 1.704 & -1.628 & -0.503 & 70 $\pm$ 60 & -5 $\pm$ 80 & 1 \nl 
S1-23 & -	& 11.9 & 1.718 & -0.900 & -1.463 & -20 $\pm$ 80 & 10 $\pm$ 90 & 2 \nl 
S1-24 & -	& 11.6 & 1.729 & 0.759 & -1.554 & -160 $\pm$ 40 & -210 $\pm$ 70 & 2 \nl 
S1-25 & -	& 13.8 & 1.810 & 1.688 & -0.655 & -70 $\pm$ 90 & 240 $\pm$ 50 & 3 \nl 
S2-1 & 29 S, W1	& 11.4 & 2.024 & -1.801 & 0.925 & -40 $\pm$ 50 & 50 $\pm$ 50 & 1 \nl 
S2-2 & -	& 14.2 & 2.068 & -0.536 & 1.998 & 140 $\pm$ 140 & 200 $\pm$ 40 & 3 \nl 
S2-3 & -	& 14.5 & 2.081 & -1.513 & -1.430 & 110 $\pm$ 110 & 260 $\pm$ 230 & 3 \nl 
S2-4 & -	& 12.1 & 2.085 & 1.457 & -1.491 & 140 $\pm$ 70 & 230 $\pm$ 40 & 3 \nl 
S2-5 & -	& 13.3 & 2.090 & 1.905 & -0.859 & 240 $\pm$ 110 & 80 $\pm$ 80 & 3 \nl 
S2-6 & -	& 11.9 & 2.097 & 1.598 & -1.358 & 200 $\pm$ 60 & 230 $\pm$ 50 & 3 \nl 
S2-7 & -	& 13.6 & 2.099 & 1.056 & 1.814 & -60 $\pm$ 90 & 220 $\pm$ 80 & 3 \nl 
S2-8 & W2	& 12.3 & 2.108 & -1.940 & 0.824 & 150 $\pm$ 110 & 120 $\pm$ 90 & 1 \nl 
S2-9 & 16 CC	& 10.6 & 2.117 & 2.055 & 0.508 & -120 $\pm$ 50 & 170 $\pm$ 60 & 2 \nl 
S2-10 & 29 N	& 9.9 & 2.125 & -1.587 & 1.413 & 200 $\pm$ 100 & -140 $\pm$ 100 & 2 \nl 
S2-11 & -	& 11.9 & 2.128 & 2.034 & -0.625 & -130 $\pm$ 80 & 110 $\pm$ 70 & 2 \nl 
S2-12 & -	& 15.8 & 2.129 & 1.759 & 1.200 & -220 $\pm$ 70 & -340 $\pm$ 170 & 3 \nl 
S2-13 & -	& 11.3 & 2.164 & -0.015 & -2.164 & -90 $\pm$ 140 & -100 $\pm$ 90 & 2 \nl 
S2-14 & -	& 15.5 & 2.170 & -1.501 & -1.567 & 180 $\pm$ 350 & 340 $\pm$ 270 & 3 \nl 
S2-15 & 16 SE1	& 10.8 & 2.195 & 1.861 & -1.164 & 170 $\pm$ 60 & 240 $\pm$ 40 & 2 \nl 
S2-16 & -	& 11.8 & 2.232 & -0.911 & 2.038 & -200 $\pm$ 90 & 10 $\pm$ 40 & 2 \nl 
S2-17 & -	& 10.9 & 2.259 & 1.270 & -1.868 & 260 $\pm$ 60 & 180 $\pm$ 40 & 2 \nl 
S2-18 & -	& 13.2 & 2.317 & -0.901 & -2.135 & -320 $\pm$ 230 & 140 $\pm$ 140 & 2 \nl 
S2-19 & -	& 12.6 & 2.334 & 0.527 & 2.273 & -180 $\pm$ 180 & 90 $\pm$ 40 & 3 \nl 
S2-20 & -	& 16.4 & 2.349 & 1.711 & 1.609 & 50 $\pm$ 200 & 340 $\pm$ 360 & 3 \nl 
S2-21 & -	& 13.6 & 2.363 & -1.696 & -1.645 & 420 $\pm$ 110 & 80 $\pm$ 100 & 1 \nl 
S2-22 & -	& 13.0 & 2.389 & 2.371 & -0.287 & -140 $\pm$ 60 & 330 $\pm$ 40 & 1 \nl 
S2-23 & -	& 14.9 & 2.460 & 1.700 & 1.779 & 240 $\pm$ 130 & -100 $\pm$ 170 & 3 \nl 
S2-24 & W16	& 13.9 & 2.464 & -2.281 & -0.934 & -140 $\pm$ 220 & 110 $\pm$ 230 & 1 \nl 
S2-25 & -	& 14.1 & 2.587 & 0.790 & -2.463 & -200 $\pm$ 290 & 200 $\pm$ 200 & 3 \nl 
S2-26 & -	& 13.5 & 2.603 & 0.732 & 2.497 & 340 $\pm$ 120 & -420 $\pm$ 100 & 3 \nl 
S2-27 & -	& 13.0 & 2.755 & 0.269 & 2.742 & -170 $\pm$ 270 & -690 $\pm$ 190 & 3 \nl 
S2-28 & -	& 14.5 & 2.847 & 2.697 & -0.913 & -170 $\pm$ 110 & 250 $\pm$ 170 & 3 \nl 
S2-29 & -	& 15.4 & 2.912 & 2.004 & -2.113 & -240 $\pm$ 310 & 40 $\pm$ 350 & 3 \nl 
S2-30 & -	& 15.3 & 2.913 & 2.913 & -0.042 & 220 $\pm$ 200 & 30 $\pm$ 240 & 3 \nl 
S2-31 & -	& 13.0 & 2.918 & 2.911 & -0.195 & -360 $\pm$ 60 & 140 $\pm$ 50 & 3 \nl 
S2-32 & -	& 12.1 & 2.981 & 1.148 & 2.751 & 120 $\pm$ 210 & 260 $\pm$ 110 & 2 \nl 
S3-1 & 16 NE & 9.0 & 3.094 & 2.892 & 1.100 & 160 $\pm$ 90 & -290 $\pm$ 30 & 2 \nl 
S3-2 & -	& 12.1 & 3.116 & 3.066 & 0.558 & 210 $\pm$ 90 & 80 $\pm$ 70 & 3 \nl 
S3-3 & -	& 15.0 & 3.177 & 3.100 & -0.695 & 110 $\pm$ 300 & 190 $\pm$ 470 & 3 \nl 
S3-4 & -	& 14.4 & 3.203 & 3.157 & -0.543 & 50 $\pm$ 140 & 240 $\pm$ 150 & 3 \nl 
S3-5 & 16 SE2	& 12.0 & 3.205 & 2.974 & -1.196 & 50 $\pm$ 90 & 260 $\pm$ 130 & 2 \nl 
S3-6 & -	& 12.8 & 3.260 & 3.259 & 0.080 & 90 $\pm$ 100 & -280 $\pm$ 70 & 3 \nl 
S3-7 & -	& 14.1 & 3.283 & 2.014 & -2.592 & -90 $\pm$ 260 & 200 $\pm$ 240 & 3 \nl 
S3-8 & -	& 14.3 & 3.495 & 3.463 & -0.471 & -90 $\pm$ 180 & 10 $\pm$ 170 & 3 \nl 
S3-9 & -	& 12.4 & 3.510 & 3.300 & -1.195 & -80 $\pm$ 70 & 20 $\pm$ 60 & 3 \nl 
S3-10 & -	& 12.0 & 3.586 & 3.395 & -1.156 & -60 $\pm$ 70 & -10 $\pm$ 120 & 3 \nl 
S3-11 & -	& 14.8 & 3.590 & 3.034 & -1.919 & -120 $\pm$ 200 & 240 $\pm$ 130 & 3 \nl 
S3-12 & 21	& 11.8 & 3.695 & 2.484 & -2.736 & -270 $\pm$ 170 & 170 $\pm$ 110 & 2 \nl 
S3-13 & -	& 13.6 & 3.939 & 3.840 & 0.875 & 210 $\pm$ 100 & 0 $\pm$ 310 & 3 \nl 
S4-1 & -	& 13.7 & 4.130 & 4.116 & -0.331 & 0 $\pm$ 210 & -320 $\pm$ 130 & 3 \nl 
S4-2 & -	& 12.7 & 4.144 & 3.779 & 1.702 & -160 $\pm$ 100 & -360 $\pm$ 80 & 3 \nl 
S4-3 & -	& 13.5 & 4.210 & 4.208 & 0.131 & 340 $\pm$ 280 & -140 $\pm$ 100 & 3 \nl 
S4-4 & -	& 12.1 & 4.320 & 3.647 & -2.317 & -100 $\pm$ 130 & -110 $\pm$ 170 & 3 \nl 
S4-5 & -	& 12.0 & 4.339 & 3.710 & -2.249 & -160 $\pm$ 100 & -150 $\pm$ 130 & 3 \nl 
\enddata
\tablenotetext{a}{
The naming convention adopted is designed to directly convey relevant 
information about the location of the source relative to the position of
SgrA$^*$.  Adopting the SgrA$^*$ position given by Menton et al. (1997),
which is accurate to 0.03 arcseconds, we divide the surrounding field
into concentric arsecond-wide annuli centered on SgrA$^*$.  Stars lying
in the central circle, which has radius 1 arcsecond, are given the names
S0-1, S0-2, S0-3, etc.  Stars lying in the annulus between radii of 1 to 2 
arcseconds are given the names S1-1, S1-2, and so on.  The number 
immediately following "S" thus refers to the inner radius of the annulus 
in which the star lies. The number following the hyphen is ordered in the sense 
of increasing distance from SgrA* within each annulus.  Of course there are 
more stars in each annulus than in our proper motion sample listed here.
When this nomenclature is applied to future 
lists of new stars, the number following the hyphen will continue to
be incremented within each annulus, ordered within each new list of added 
stars in the sense of increasing distance from SgrA$^*$}
\tablenotetext{b}{The epoch of the brightness and positional measurement
is 1995.4.}
\tablenotetext{c}{The brightness of each star is assessed by 
carrying out aperture photometry on the "cores" with an the object radius 
extending from 0 to 0\farcs 06, a sky annulus ranging from 0\farcs 06 to 0\farcs 09,
and a zero point determined by IRS 16 NE which is assumed 
to have a K magnitude of 9.0. } 
\tablecomments{1 - additional measurements of these stars can be found in 
Genzel et al. (1997);  2 - additional measurements of these stars can be found
in Eckart \& Genzel (1997); 3 - this is the first measurement of these stars'
proper motions}
\end{deluxetable}

{\centering \leavevmode
\epsfxsize=0.45\columnwidth \epsfbox{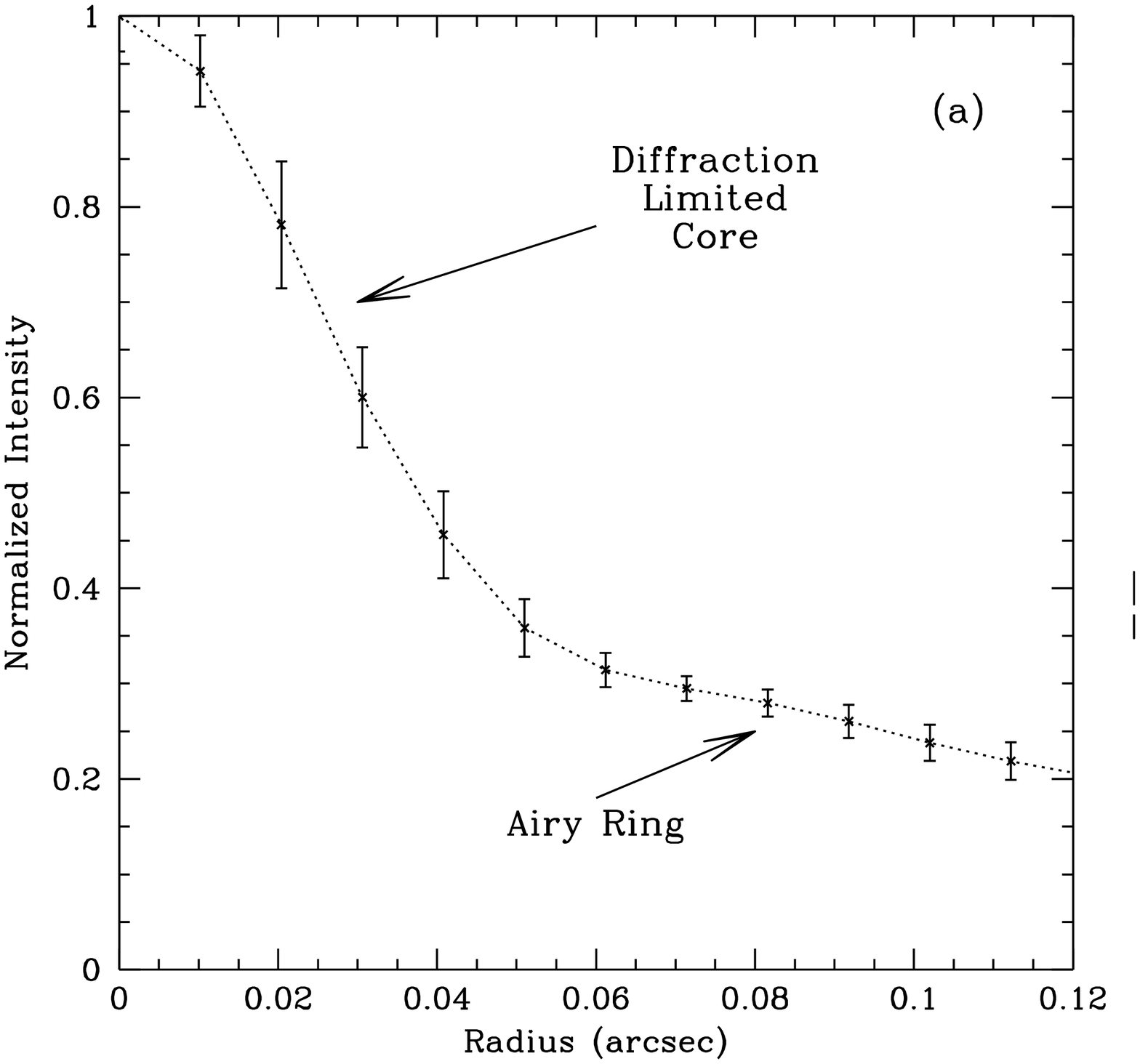} 
\hspace*{0.10\columnwidth} 
\epsfxsize=0.45\columnwidth \epsfbox{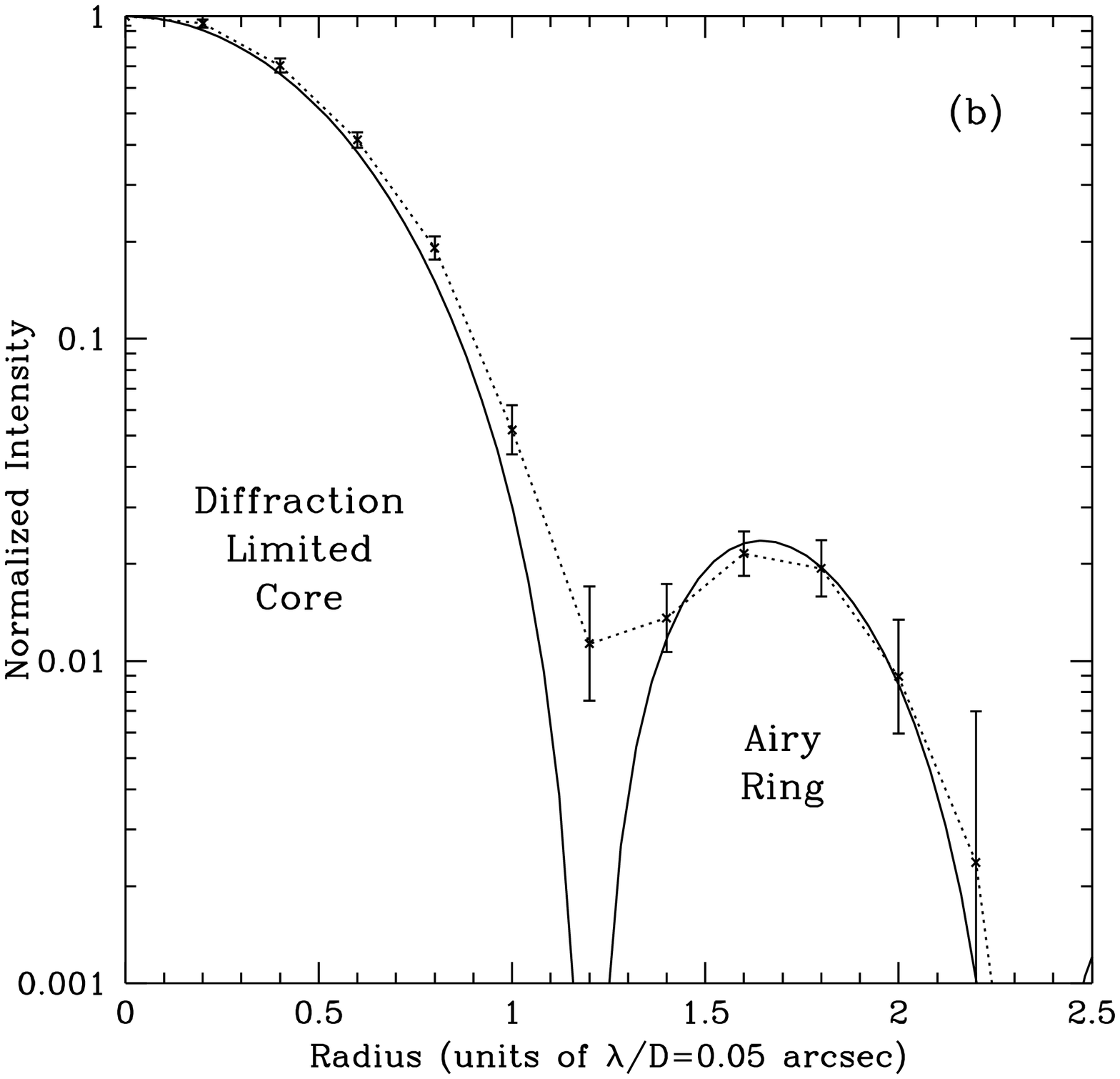} 

Figure 2}

{\centering \leavevmode
\epsfxsize=1.00\columnwidth \epsfbox{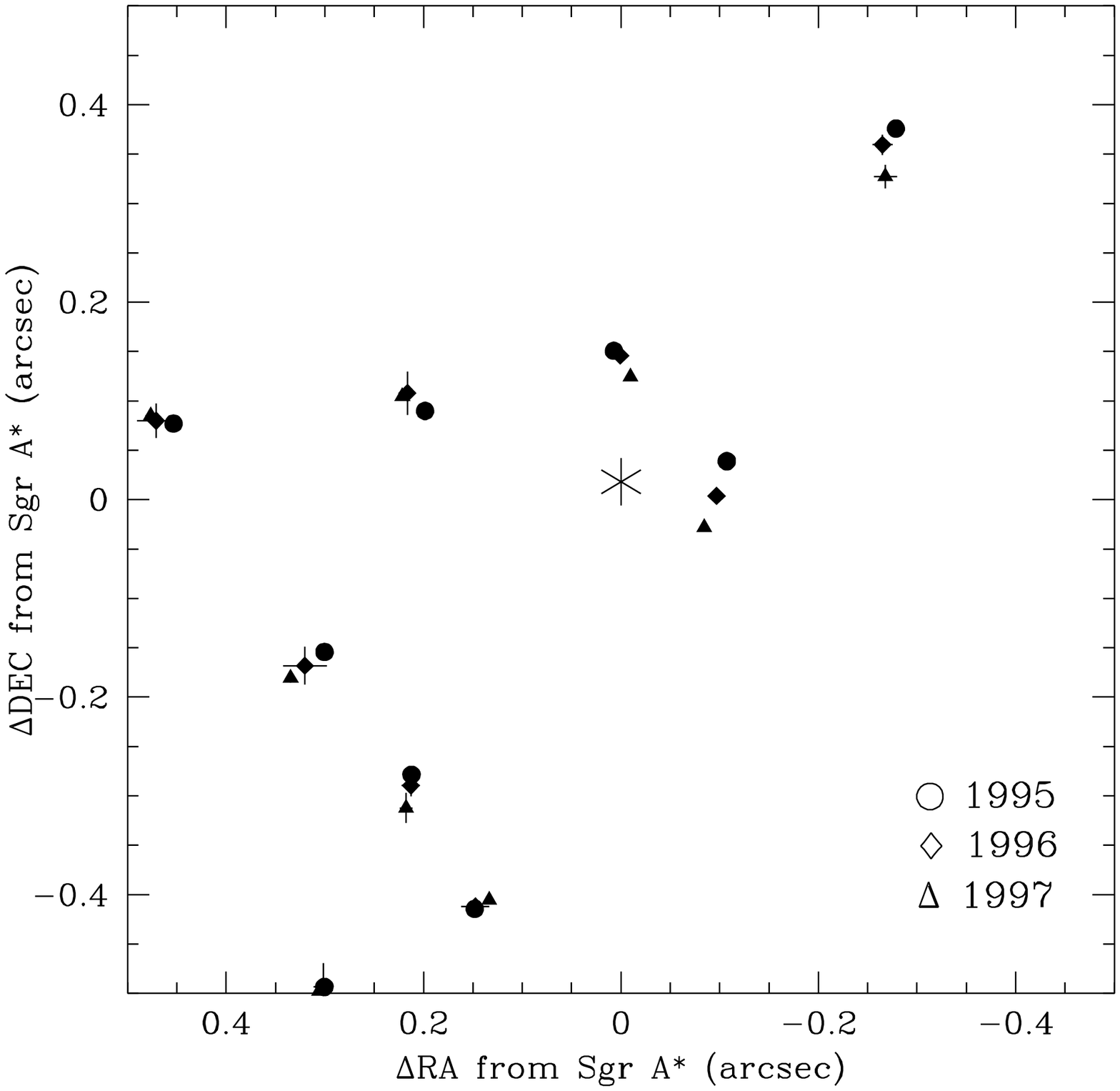} 

Figure 3}

{\centering \leavevmode
\epsfxsize=1.00\columnwidth \epsfbox{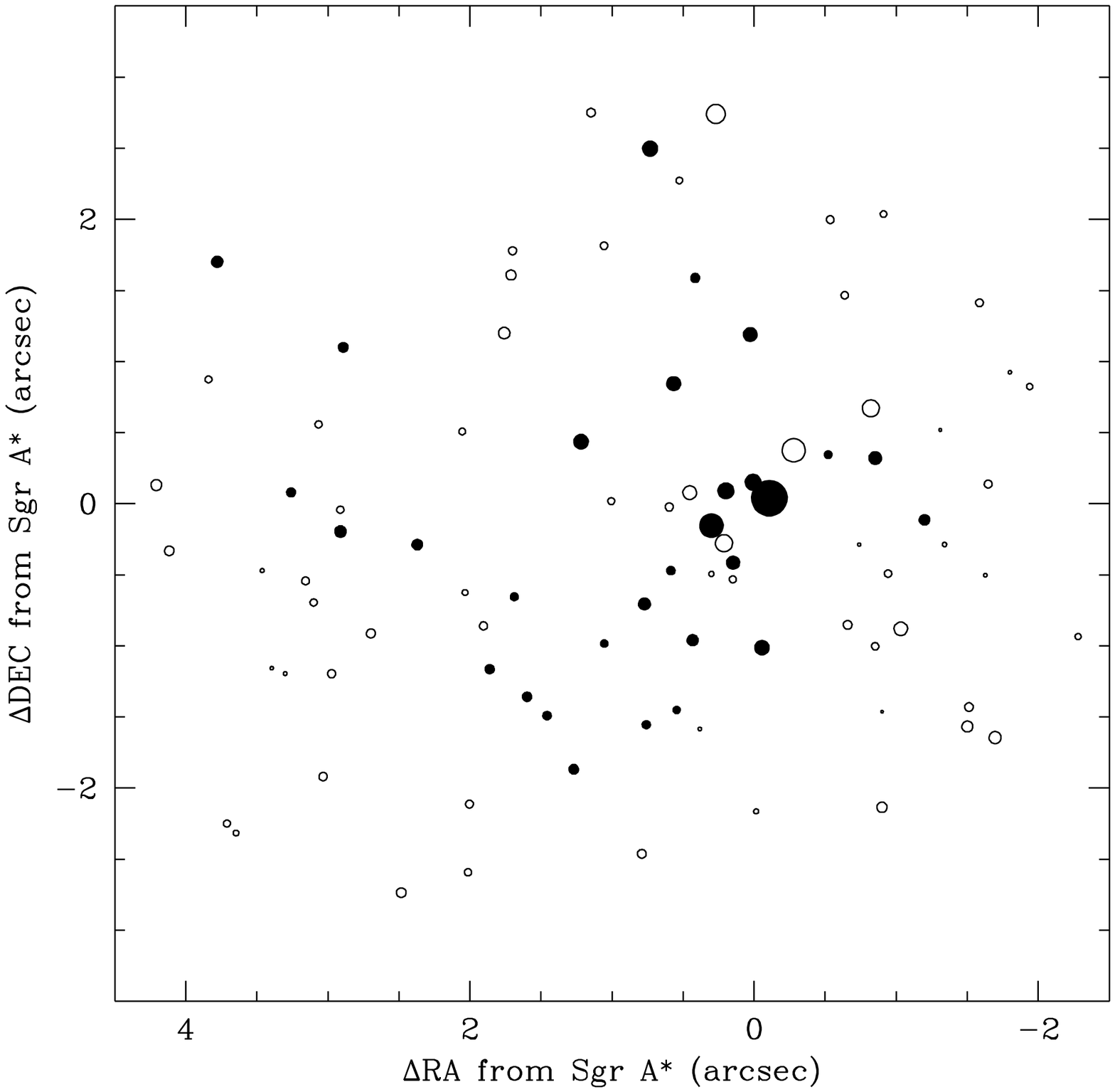} 

Figure 4}

{\centering \leavevmode
\epsfxsize=1.00\columnwidth \epsfbox{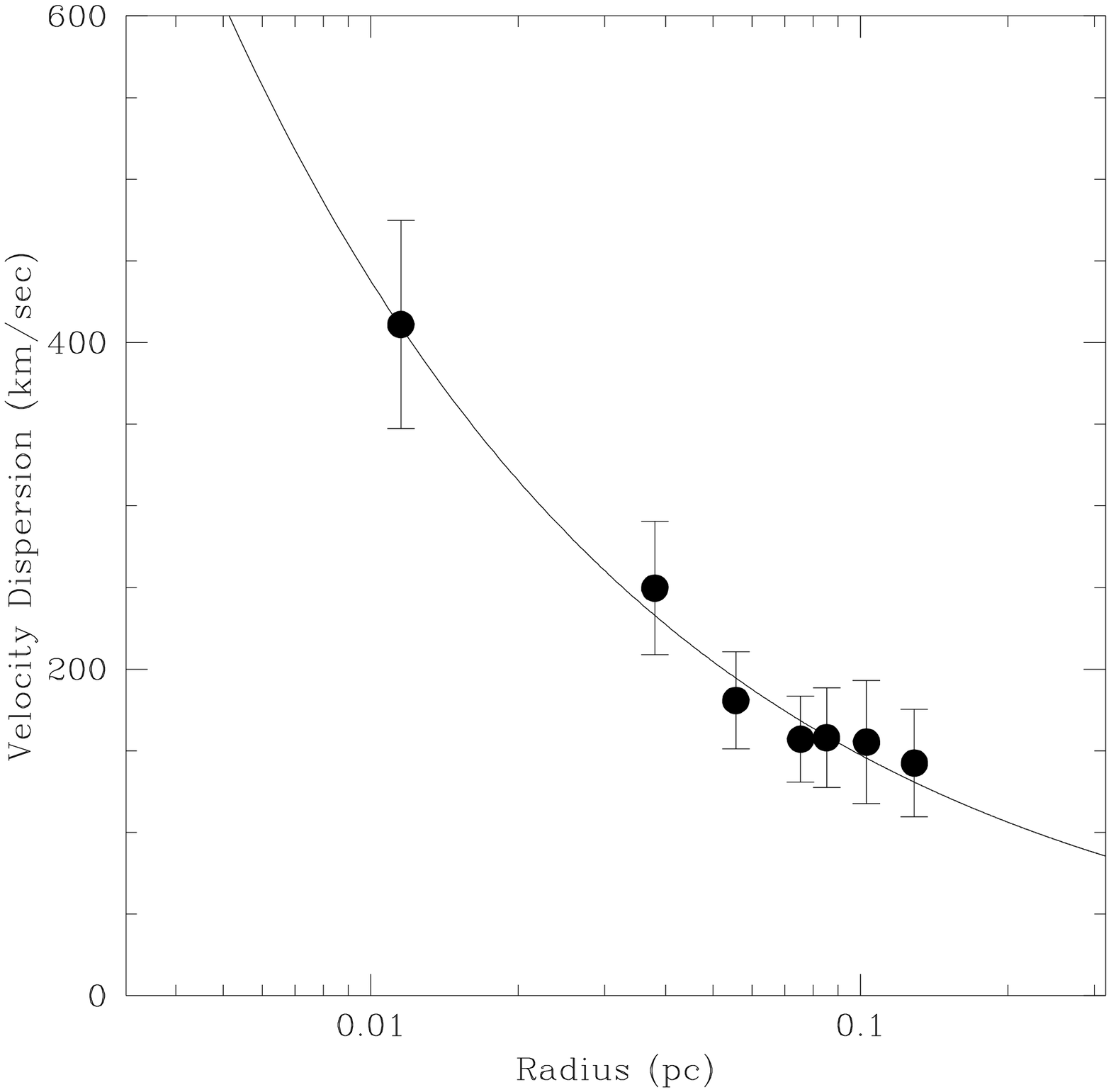} 

Figure 5}

{\centering \leavevmode
\epsfxsize=1.00\columnwidth \epsfbox{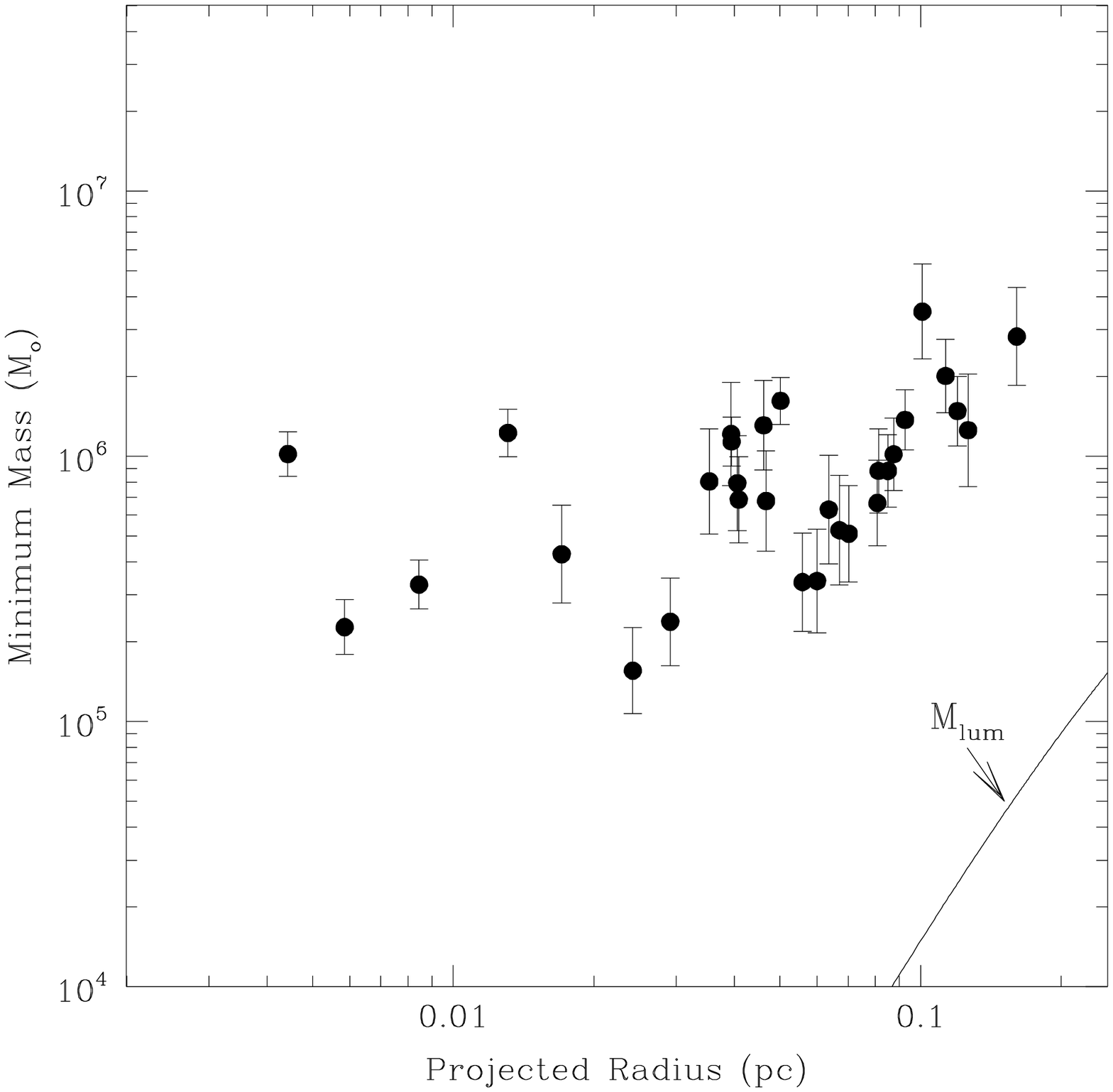} 

Figure 6}

\end{document}